\documentclass[14pt,preprint2,longabstract]{aastex}
\shorttitle{A {\it Chandra} survey of Galactic globular clusters}
\shortauthors{Cheng et al.}

\begin{document}
\title{A {\it Chandra} Survey of Milky Way Globular Clusters II: Testing the Hills-Heggie Law}

\author{Zhongqun Cheng$^{1,2,3}$, Zhiyuan Li$^{2,3}$, Xiaojie Xu$^{2,3}$, Xiangdong Li$^{2,3}$ Zhenlin Zhu$^{2,3}$ and Taotao Fang$^{1}$}
\affil{$^{1}$ Department of Astronomy and Institute for Theoretical Physics and Astrophysics, Xiamen University, Xiamen, Fujian 361005, China} 
\affil{$^{2}$ School of Astronomy and Space Science, Nanjing University, Nanjing 210023, China} 
\affil{$^{3}$ Key Laboratory of Modern Astronomy and Astrophysics (Nanjing University), Ministry of Education, Nanjing 210023, China}

\email{lizy@nju.edu.cn; lixd@nju.edu.cn; fangt@xmu.edu.cn}

\begin{abstract}
Binary-single and binary-binary encounters play a pivotal role in the evolution of star clusters, as they may lead to the disruption or hardening of binaries, a novel prediction of the Hills-Heggie law. 
Based on our recent {\it Chandra} survey of Galactic globular clusters (GCs), we revisit the role of stellar dynamical interactions in GCs, focusing on main-sequence (MS) binary encounters as a potential formation channel of the observed X-ray sources in GCs.
We show that the cumulative X-ray luminosity ($L_X$), a proxy of the total number of X-ray-emitting binaries (primarily cataclysmic variables and coronally active binaries) in a given GC, is highly correlated with the MS binary encounter rate ($\Gamma_{b}$), as $L_X \propto \Gamma_{b}^{0.77\pm0.11}$. 
We further test the Hills-Heggie law against the {\it binary hardness ratio}, defined as the relative number of X-ray-emitting hard binaries to MS binaries and approximated by $L_{X}/(L_{K}f_{b})$, with $L_{K}$ being the GC K-band luminosity and $f_b$ the MS binary fraction. We demonstrate that the binary hardness ratio of most GCs is larger than that of the Solar neighborbood stars, and exhibits a positive correlation with the cluster specific encounter rate ($\gamma$), as $L_{X}/(L_{K}f_{b}) \propto \gamma^{0.65\pm0.12}$.  
We also find a strong correlation between the binary hardness ratio and cluster velocity dispersion ($\sigma$), with $L_{X}/(L_{K}f_{b}) \propto \sigma^{1.71\pm0.48}$, which is consistent with the Hills-Heggie law.  
We discuss the role of binary encounters in the context of the Nuclear Star Cluster, arguing that the X-ray-emitting, close binaries detected therein could have been predominatly formed in GCs that later inspiralled to the Galactic center. 

\end{abstract}

\section{Introduction}

X-ray observations in the 1970s have discovered a strong over-abundance of outbursting low-mass X-ray binaries (LMXBs) in globular clusters (GCs) with respect to the Galactic field \citep{clark1975,katz1975}, which quickly stimulated the discussions of stellar dynamical interactions as an effective formation channel of LMXBs in GCs. Widely accepted scenarios include: tidal capture of neutron star (NS) by main sequence (MS) stars \citep{fabian1975}, collision of NS with giant stars \citep{sutantyo1975}, and exchange encounter of NS with primordial binaries \citep{hills1976}. All these mechanisms have a similar dependence on the rate of stellar close encounters in GCs, which can be expressed as $\Gamma \propto \int {\rho}^{2}/\sigma$, with $\rho$ the stellar density and $\sigma$ the velocity dispersion. Given $\Gamma$, one can readily predict the probability of finding LMXBs in GCs of the Milky Way or external galaxies \citep{verbunt1987,jordan2004,sivakoff2007,jordan2007,peacock2009}. 
 
Studies of stellar dynamical interactions in the 1970s and 1980s also predicted a possible formation mechanism of close binaries in GCs, namely, the binary-single ({\it b-s}) and binary-binary ({\it b-b}) encounters \citep{hills1975,heggie1975,hoffer1983,mikkola1983,mikkola1984a, mikkola1984b, hut1992a,hut1992b,hut1993,bacon1996}. Due to energy equipartition, stars involved in these interactions tend to have the same kinetic energy, which will change the bound energy of the binary and lead to expansion or contraction of the binary orbit. Generally, the evolution of binary orbits in dynamical encounters are dictated by the Hills-Heggie law: hard binaries (with bound energy $|E_{b}|$ larger than the kinetic energy $E_{k}$ of the intruding star) tend to be harder, while soft binaries ($|E_{b}| < E_{k}$) tend to be softer and eventually be disrupted \citep{hills1975,heggie1975,hut1993}. 

For the numerous MS binaries in GCs, the Hills-Heggie law provides an effective mechanism for transforming them into exotic binaries, which has been confirmed by numerical simulations. For example, the exchange encounters of MS binaries with compact objects may lead to the formation of cataclysmic variables (CVs) or LMXBs \citep{heggie1996, rasio2000, ivanova2006, ivanova2008}. In the case of no exchage, the MS binaries of short orbital periods may evolve toward tidal locking between stellar rotation and orbital motion, effectively enhancing the stellar magnetic activities and transforming them into coronally active binaries (ABs). Meanwhile, enhancing the rate of binary mass transfer or merger can lead to the formation of blue stragglers \citep{hut1992a, fregeau2004, chatterjee2013}.

On the observational side, the {\it Chandra X-ray Observatory} has revealed numerous weak X-ray sources (with luminosities $\lesssim 10^{34}{\rm~erg~s^{-1}}$) in GCs, the majority of which are found to be CVs and ABs, with a small addition of quiescent LMXBs and millisecond pulsars (MSPs) (see \citealp{heinke2010} for a recent review). All these systems are close binaries. \citet{pooley2003} were among the first to suggest that these sources are dynamically originated, based on the observed correlation between the number of detected X-ray sources ($N_{X}$) and $\Gamma$ in a sample of Galactic GCs. It is noteworthy that the $N_{X}-\Gamma$ correlation was found to be sub-linear\footnote{The logarithmic slope of the $N_{X}-\Gamma$ relation in \citet{pooley2003} is $0.74\pm0.36$, which has been revised as $0.55\pm0.09$ in \citet{maxwell2012}.}, which suggests a lower formation efficiency of weak X-ray sources in more massive GCs (with higher $\Gamma$). 

Based on archival {\it Chandra} observations, we have recently carried out an X-ray survey of 69 Galactic GCs \citep[hereafter Paper I]{cheng2018}. Using the cumulative X-ray luminosity ($L_{X}$) as a proxy of the weak X-ray source populations, we have shown that $L_{X}$ is highly correlated with not only $\Gamma$, but also the cluster mass ($M$), suggesting that the primordial channel of close binary formation is not negligible. In the meantime, our larger GC sample disapproves a positive correlation between the source abundance (approximated by $L_{X}/L_{K}$, where $L_{K}$ is the GC K-band luminosity, itself a good proxy of the cluster mass; Paper I) and the specific encounter rate, ${\gamma}={\Gamma}/M$, which was originally suggested by \citet{pooley2006}. Furthermore, we demonstrated that the weak X-ray sources (mainly CVs and ABs) are under-abundant in GCs with respect to the field, a behavior opposite to the LMXBs and MSPs (see also \citet{xu2018} for the case of M31 GCs). These findings render the dynamical origin of GC weak X-ray sources far from conclusive.
   
The problem may lie in the ambiguity of the stellar dynamical interactions collectively expressed by the parameter $\Gamma$. In particular, when considering the formation of close binaries in a dense stellar environment such as GCs, there exist two competing dynamical processes: single-single ({\it s-s}) encounters (tidal capture, direct collision, etc.) that usually lead to the dissipation of stellar kinetic energy and result in the formation of binaries, while {\it b-s} and {\it b-b} encounters tend to modify or destroy binaries, leading to a gradual net decrease of binary abundance \citep{hills1975,heggie1975,mikkola1983,mikkola1984a, mikkola1984b, hut1992a,hut1992b,hut1993}. For CVs and ABs, the binary-relevant encounters are non-negligible, because their progenitor MS binaries are mainly low-mass stars, which will evolve on a time-scale comparable to or even greater than the GC relaxation time. 

According to the Hills-Heggie law, the evolution of MS binaries in GCs depends on the average kinetic energy $E_{k}$ of the intruding star, which is related to the stellar velocity dispersion as $E_{k}\sim \sigma^{2}$. There exists a watershed orbital separation ($a_w$) for the MS binaries, with $E_{b} = GM_{\ast}^{2}/2a_{w} \sim E_{k}$, and hence $a_{w}\propto \sigma^{-2}$. For MS binaries with an orbital separation greater than $a_w$, they are more likely to be disrupted by {\it b-s} encounters, and to a lesser extent, by {\it b-b} encounters (see below). Otherwise, the MS binaries will be dynamically transformed into close binaries and ultimately become ABs or CVs. 
The prediction of dynamical disruption of binaries has been confirmed by the observed anti-correlation between MS binary fraction ($f_{b}$) and $\sigma$ in GCs \citep{milone2012}, and is also supported by the abnormally low values of $f_{b}$ in the core of some clusters \citep{degrijs2013}. 

On the other hand, the effect of dynamical hardening of binaries in GCs, as also predicted by the Hills-Heggiew law, remains to be tested. We aim to provide a statistical test of this effect in the present work. 
In Section 2, we first introduce the binary-relevant encounter rate ($\Gamma_{b}$), and estimate their importance relative to the single-single encounter rate ($\Gamma_{s}$) in GCs. In Section 3, we explore the formation of weak X-ray sources in GCs according to the binary encounter scenratio. In Section 4, we test the Hills-Heggie law in the scope of binary hardening in GCs. In Section 5, we discuss the role of binary encounters in the context of X-ray sources in the Nuclear Star Cluster. Concluding remarks are provided in Section 6.
  
\section{Binary-single encounter rate} 

To compare the two competing dynamical processes in GCs, we estimate the {\it s-s} and {\it b-s}/{\it b-b}\footnote{Here we make no distinction between the {\it b-b} and {\it b-s} encounters, since both of them obey the Hill-Heggie law and have a similar effect on modifying binaries \citep{mikkola1983, mikkola1984a, mikkola1984b, bacon1996}. See also Footnote~5 below.} encounter rates separately, which are defined as an integration over the cluster volume \citep{verbunt2003}:
\begin{equation}
\Gamma_{s} \propto \int n_{s}^{2}A_{s}v dV \propto \int \frac{n_{s}^{2}R_{\ast}}{v} dV,
\end{equation}
\begin{equation}
\Gamma_{b} \propto \int {n_{b} n A_{b} v dV} \propto \int {\frac{n_{b} n a}{v} dV},
\end{equation}
where $n_{s}$ ($n_{b}$) is the number density of single stars (MS binaries), $n=n_{s}+n_{b}$, and $A_{s}$ ($A_{b}$) the encounter cross-section, which is proportional to the stellar radius $R_{\ast}$ (binary orbital separation $a$) and inversely proportional to the square of relative velocity $v$ \citep{davies2002}. Generally, the relative velocity $v$ can be characterized by the stellar velocity dispersion $\sigma$. The density of MS binaries ($n_{b}$) can be related to the observed binary faction $f_{b}$, with $f_{b}=n_{b}/n$, while the stellar density $n$ can be approximated by the stellar luminosity density, $\rho=n\langle L \rangle$, where $\langle L \rangle$ is the characteristic luminosity of MS stars. Now, Equations (1) and (2) can be rewritten as:
\begin{equation}
\Gamma_{s} \propto \int \frac{(1-f_{b})^{2} \rho^{2}R_{\ast}}{\sigma}dV,
\end{equation}
\begin{equation}
\Gamma_{b} \propto \int \frac{f_{b} \rho^{2} a}{\sigma}dV.
\end{equation}

To accurately calculate $\Gamma_{s}$ and $\Gamma_{b}$, a robust measurement of the radial stellar density profile $\rho(r)$ and binary fraction profile $f_{b}(r)$ is needed.
However, due to the low intrinsic luminosities and crowded stellar environment, the identification of MS binaries in GCs is challenging. Consequently, accurate measurements of $f_{b}$ and $a$ as a function of distance from the cluster center are currently absent \citep{sollima2007,milone2012,ji2015}.
Therefore, we estimate $\Gamma_{s}$ and $\Gamma_{b}$ by considering a global $f_{b}$ for a given GC, and take $R_{\ast}$ and $a$ as constants throughout the cluster\footnote{The radii of low mass MS stars should have a narrow range, but the size of binaries may vary by orders of magnitude, which will lead to an overestimate of $\Gamma_{b}$ in core-collapsed GCs. See Section 3 for further discussions.}. Equations (3) and (4) can then be further simplified as $\Gamma_{s} \propto (1-f_{b})^{2} \times \Gamma$ and $\Gamma_{b} \propto f_{b} \times \Gamma$, respectively. 
By choice, both $f_{b}$ and $\Gamma$ are global parameters and are more compatible with the GC cumulative X-ray luminosity as measured in Paper I. 

To find out the dominant type of encounters\footnote{The ratio of the {\it b-s} and {\it b-b} encounters can be estimated with ${\Gamma_{bs}}/{\Gamma_{bb}}\sim (n_{b}n_{s}A_{b})/(n_{b}n_{b}A_{b}) \sim (1-f_{b})/f_{b}$. The {\it b-s} encounters dominate over {\it b-b} encounters in GCs: ${\Gamma_{bs}}/{\Gamma_{bb}}$ ranges from 4 to 99 when $f_{b}$ ranges from $1\%$ to $20\%$ \citep{milone2012}.}, we estimate the ratio of the two competing dynamical interactions with Equations (3) and (4):
\begin{equation}
\frac{\Gamma_{b}}{\Gamma_{s}} \sim \frac{f_{b}}{(1-f_{b})^{2}} \frac{a}{R_{\ast}}.
\end{equation}
Considering a distribution of MS binary orbital period ranging from $\sim10-5000$ days\footnote{According to Equation (1) of \citet{hut1992b}, binaries with orbital period less than $\sim10$ days may suffer from no strong encounters in typical GCs with $\rho_{c}\sim10^{4}\rm\, M_{\odot}\,pc^{-3}$ and age of $\sim10 \rm \, Gyr$. On the other hand, for a GC with stellar velocity dispersion $\sigma\sim10 \rm \, km\, s^{-1}$, the watershed orbital separation $a_{w}$ corresponds to a watershed orbital period $\sim5000$ days, provided that the constituent stars have masses of $\sim0.8{\rm~M_{\odot}}$.}, we have values of $a/R_{\ast}$ ranging from $30$ to $1900$ for a typical MS mass of $0.8{\rm~M_{\odot}}$. The observed binary fraction $f_{b}$ in most GCs ranges from $1\%$ to $20\%$ \citep{milone2012}, thus the ratio ${\Gamma_{b}}/{\Gamma_{s}}$ ranges from $0.3$ to $600$. 
Since most MS binaries have been dynamically exhausted in GCs and the present-day $f_{b}$ is smaller than it used to be, the ratio of ${\Gamma_{b}}/{\Gamma_{s}}$ should be even larger in the past. 
Moreover, due to mass segregation, MS binaries are more likely to sink to the cluster core and thus subject to higher encounter rates than single stars. Such an effect has been ignored in Equation (5), but otherwise would enhance the value of ${\Gamma_{b}}/{\Gamma_{s}}$. 
Therefore, the {\it b-s}/{\it b-b} encounters dominate over the {\it s-s} encounters in GCs. In the following, we will focus on the binary-relevant encounters when considering the formation of weak X-ray sources.

\section{Formation of hard X-ray binaries by binary-single encounters}

Among the 69 GCs studied in Paper I, thirty, including 6 core-collapsed GCs and 24 dynamically normal GCs, have a measured MS binary fraction ($f_b$) from \citet{milone2012}.  
Using the method outlined in Section 2, we calculate the MS binary encounter rate for each of these 30 GCs. 
We adopt the total encounter rate by \citet{bahramian2013}, which was calculated as $\Gamma=4\pi\sigma_{c}^{-1}\int\rho^{2}(r)r^{2}dr$\footnote{Here the distribution of $\sigma(r)$ was assumed to be flat, with a value equivalent to the cluster central velocity dispersion $\sigma_{c}$. Such a simplification is reasonable, since the profile of $\sigma(r)$ is much flatter than $\rho(r)$ in GCs.} and normalized to a reference value of $1000$ for NGC\,104.
Thus the binary-related encounter rate of NGC\,104 is $\Gamma_b=180^{+66}_{-65}$, for its MS binary fraction $f_b = (1.8\pm0.6)\%$.
The relative errors of $f_{b}$ and $\Gamma$ in most GCs are $\lesssim$30\% (Table 1), which have been included in the error budget of $\Gamma_{b}$ through standard error propagation.

\begin{figure*}[ht]
\centering
\begin{minipage}[!htbp]{1.0\textwidth}
\leftline{\includegraphics[angle=0,origin=br,height=0.3\textheight, width=0.5\textwidth]{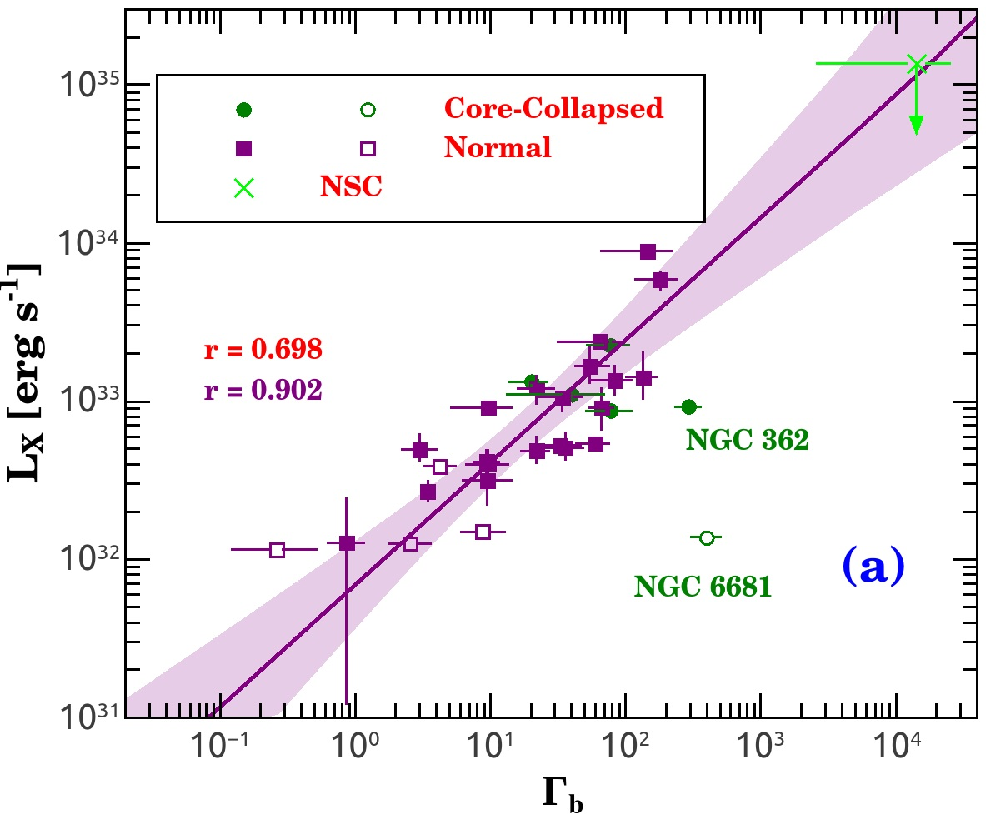}
\includegraphics[angle=0,origin=br,height=0.3\textheight, width=0.5\textwidth]{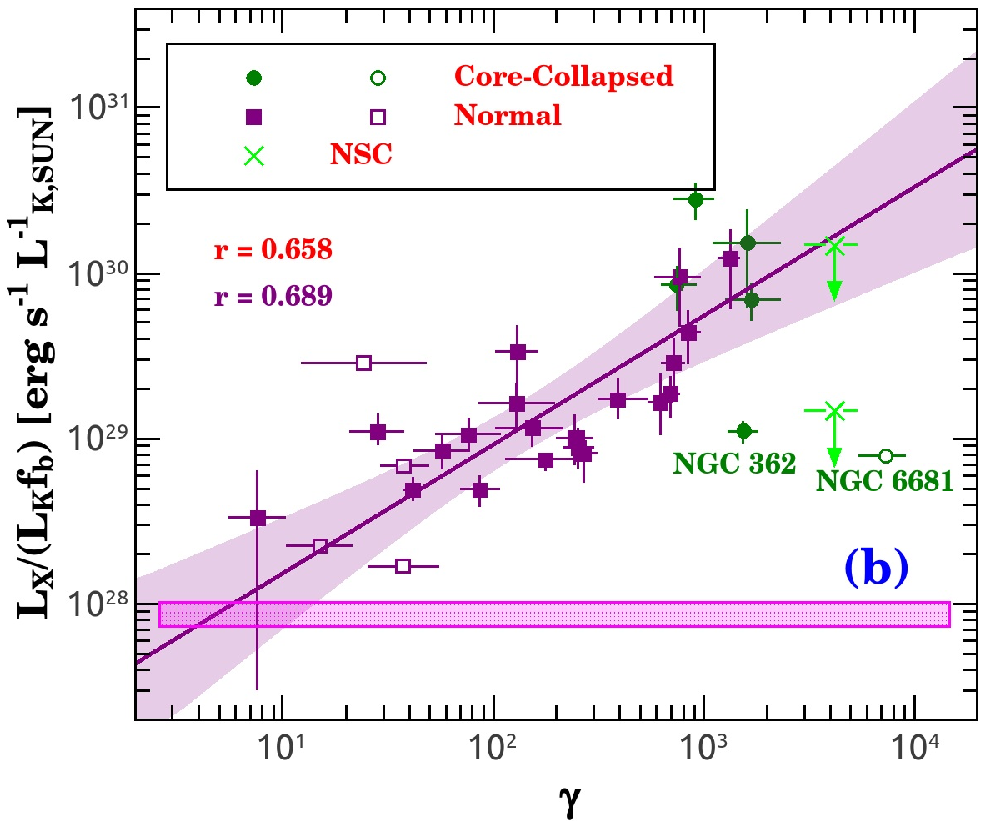}}
\linespread{0.7}
\caption{(a): GC X-ray luminosity as a function of the MS binary encounter rate; (b): $L_{X}/(L_{K}f_{b})$ as a function of the specific encounter rate $\gamma$. The olive circles and purple squares denote the core-collapsed and dynamically normal GCs, with filled and open symbols represent the actual detection and the upper limit (at $95\%$ confidence). The green crosses mark the upper limits of thbinaries) in the Nuclear Star Cluster, and $L_{X}/(L_{K}f_{b})$ in (b) were calculated with $f_{b}=1\%$ and $f_{b}=10\%$ separatively. The purple and red text indicates the Spearman's rank correlation coefficient of the dynamically normal and total GCs, respectively. The purple solid line is the best-fitting function for dynamically normal GCs (GCs with upper limit were not included in the fit), and the shaded area represents the $95\%$ confidence interval. Magenta strip in (b) is the Solar Neighborhood stars. \label{fig:LX-Gamb}}
\end{minipage}
\end{figure*}

Figure~\ref{fig:LX-Gamb}a displays the GC X-ray luminosity versus $\Gamma_{b}$. The values of $L_{X}$ were adopted from Paper I. GCs of solid detections are shown by filled symbols (with 68\% uncertainties), while X-ray non-detected GCs are denoted by open symbols representing the 95\% upper limit.
For the dynamically normal GCs, there is a significant correlation between $L_{X}$ and $\Gamma_{b}$, with Spearman's rank correlation coefficient $r=0.902$ and random correlation p-value $p\ll0.0001$. We fit the relation with a power-law function, which is plotted as the purple line in Figure 1a, with $L_{X} \propto \Gamma_{b}^{0.77\pm0.11}$. 

We note that in Paper I, a strong correlation between $L_X$ and $\Gamma$ has also been found for the full sample of dynamically normal GCs, with $L_{X}\propto \Gamma^{0.79\pm0.12}$. Here, for the subset of dynamically normal GCs selected in Figure \ref{fig:LX-Gamb}a, the correlation between $L_X$ and $\Gamma$ is also significant, with the Spearman's correlation coefficient $r=0.896$ and random correlation p-value $p\ll0.0001$. 
However, we argue that $\Gamma_{b}$ be more fundamental than $\Gamma$ in describing the origin of the weak X-ray sources in GCs. 
First, as discussed in Section 2, the typical value of $\Gamma_{b}/\Gamma_{s}$ is much larger than unity in GCs, and the encounters evaluated in $\Gamma$ are actually dominated by $\Gamma_{b}$. 
%More importantly, the empirical relations of $N_{X}-\Gamma$ and $L_{X}-\Gamma$ have been shown to be sub-linear, 
Moreover, the sublinear relation of $L_{X}-\Gamma$ can be understood as due to binary-related encounters. 
This is because stars in more massive GCs will be more energetic ($E_{k}\propto \sigma^{2}$), thus more MS binaries will be dynamically disrupted rather than be transformed into X-ray-emitting close binaries. As a result, formation of weak X-ray sources is less efficient in massive GCs than in low-mass clusters, as evidenced by an anti-correlation between the GC X-ray emissivity and cluster mass\footnote{This anti-correlation has been confirmed by \citet{xu2018} with 44 GCs in M31.}, $\epsilon_{X}\equiv L_{X}/M \propto M^{-0.30\pm0.11}$ (Paper I). 

The core-collapsed GCs in Figure \ref{fig:LX-Gamb}a deserve some remarks. These clusters appeared to be abundant in X-ray sources according to the $N_{X}-\Gamma$ relation of \citet{pooley2003}, however, with the updated $\Gamma$, \citet{bahramian2013} found a paucity of X-ray sources in these clusters. The binary encounter scenario provides an explanation for these clusters: due to the much older dynamical age, MS binaries in core-collapsed GCs are on average much harder than those in dynamically normal GCs (Figure~\ref{hardness-sigma}). Thus, taking the binary orbital separation $a$ as a constant in Equation (4) could have led to an overestimated $\Gamma_{b}$ for these GCs (e.g., NGC\,362 and NGC\,6681 as denoted in Figure~\ref{fig:LX-Gamb}a). 

In Figure~\ref{fig:LX-Gamb}a, we also include the case of the Nuclear Star Cluster (NSC) for comparison.
The vast majority of the weak X-ray sources (with luminosities spanning $10^{31}-3\times10^{33}\rm\ erg\,s^{-1}$) detected in the NSC are thought to be CVs \citep{zhu2018}.
The cumulative X-ray luminosity of the NSC is estimated from the X-ray source catalog of \citet{zhu2018}, as follows. From the accumulated spectrum of the sources detected within a galactocentric radius $r<100\arcsec$, we derive an unabsorbed luminosity of $3.3\times 10^{34}\rm\ erg\,s^{-1}$ in 0.5-8 keV band. 
As shown in \citet{zhu2018}, these sources account for a resolved fraction of $\sim$24\%. Therefore, we adopt $1.37\times 10^{35}\rm\ erg\,s^{-1}$ as a firm upper limit of the cumulative luminosity of all weak X-ray sources in the NSC.
To estimate the encounter rate of the NSC, we follow the procedure of \citet{bahramian2013} and numerically integrate $\rho^{2}(r)/\sigma(r)$ throughout the NSC and NGC\,104, the latter serving as a norm here. We have adopted the stellar density and velocity dispersion profiles as determined by \citet{fritz2016} for the NSC, the cumulative stellar mass of which within was estimated to be $(6.09\pm0.97)\times 10^{6}\rm\ M_{\odot}$ within $r<100\arcsec$ (Table 1). For the reference value of $\Gamma=1000$ for NGC\,104, we obtain $\Gamma=25600\pm7700$ for the NSC. Since the MS binary fraction of the NSC is currently poorly constrained, we assume two plausible values of $f_{b}=1\%$ and 10\%\footnote{Such a range of $f_{b}$ is theoretically predicted by numerical simulations considering binary evolution in the NSC (see Appendix C of \citet{generozov2018} for details).}, which results in $\Gamma_{b}$ ranging from 2560 and 25600. From Figure~\ref{fig:LX-Gamb}a, it can be seen that the NSC is compatible with the relation defined by the GCs. The implication of this finding will be addressed in Section 5.

As shown in Paper I, the cumulative GC X-ray luminosity may be related to the number ($N_{X}$) of the weak X-ray sources, $L_{X}=N_{X}\overline{L}_{X}$, and the abundance of weak X-ray source can be expressed as,
\begin{equation} 
L_{X}/L_{K}=f_{b}(N_{X}/N_{b})(\overline{L}_{X}/\overline{L}_{K}),
\end{equation}
where $L_{K}$ is the GC cumulative K-band luminosity, $\overline{L}_{X}$ ($\overline{L}_{K}$) the characteristic X-ray (K-band) luminosity of a binary (star), and $N_{X}/N_{b}$ the fraction of binaries being an X-ray-emitting close binary. 
We follow Paper I to calculate $L_{X}$ and $L_{K}$ from the same photometry extraction region (i.e., within the half-light circle). Both $L_{X}$ and $L_{K}$ have been corrected from extinction with the foreground reddening ($E(B-V)$) of \citet{harris1996}. It turns out that the measurement error of $L_K$ is small (with relative error $\lesssim$1\%) compared to that in $L_{X}$ (with relative error $\lesssim$30\% in most GCs), hence the error in $L_{X}/L_{K}$ is dominated by $L_{X}$. 
According to Equation (6), we can use the parameter $L_{X}/(L_{K}f_{b})$ to diagnose the dynamical interactions of binaries in GCs, which could be regarded as the transformation rate of MS binaries into X-ray-emitting close binaries. We refer to this parameter as the {\it binary hardness ratio}. 

In Figure~\ref{fig:LX-Gamb}b, we plot $L_{X}/(L_{K}f_{b})$ versus the specific encounter rate $\gamma$, which is defined as $\gamma\equiv\Gamma/M_{6}$, with the value of $\Gamma$ adopted from \citet{bahramian2013}, and $M_{6}$ is the cluster mass in units of $10^{6}{\rm M_{\odot}}$. 
The Spearman's rank correlation coefficient shows a significant positive correlation between these two parameters, with $r=0.689$ and $r=0.658$ for the dynamically normal and total GCs, while the random correlation p-value is $p=0.0002$ and $p<0.0001$, respectively. 
%Clearly, the dynamically normal GCs with higher $\gamma$ are more likely to have a higher value of $L_{X}/(L_{K}f_{b})$. 
This clearly supports a dynamical origin of weak X-ray sources in GCs, in particular the dynamically normal ones. We fit these GCs with a power-law function, which gives $L_{X}/(L_{K}f_{b}) \propto \gamma^{0.65\pm0.12}$ (purple line in Figure~\ref{fig:LX-Gamb}b).

For comparison, we also plot the cases of the NSC and Solar neighborhood in Figure~\ref{fig:LX-Gamb}b. The NSC is marked by the green crosses, with the upper limit of $L_{X}/(L_{K}f_{b})$ varying from $1.48\times 10^{29} \rm \ erg\,s^{-1}L^{-1}_{K,\odot}$ to $14.8\times 10^{29} \rm \ erg\,s^{-1}L^{-1}_{K,\odot}$ ($f_b$ from 1\%--10\%). Here, the uncertainty in $L_{X}/(L_{K}f_{b})$ is dominated by the poorly constrained $f_{b}$ in the NSC. 
In case of the latter upper limit, the NSC is again consistent with the relation defined by the GCs in Figure~\ref{fig:LX-Gamb}b.

Due to the collisionless environment, the Solar neighborhood is marked as the magenta horizontal strip in Figure~\ref{fig:LX-Gamb}b, with $L_{X}/(L_{K}f_{b})=(8.6\pm1.6)\times 10^{27} \rm \ erg\,s^{-1}L^{-1}_{K\odot}$. Estimate of this value has adopted the X-ray emissivity of Solar neigborhood stars from \citet{sazonov2006} and \citet{revnivtsev2007}, the mass-to-light ratio for the Solar cylinder of $M/L_{K}=0.34{\rm~M_{\odot}}/L_{K\odot}$ \citep{just2015}, and the frequency of Solar neighborhood stars in binary or multiple systems of $46\pm2\%$ \citep{raghavan2010}. 
If all the primordial binaries in GCs followed the normal stellar evolution path as in the Galactic field binaries, the GCs should have a similar value of $L_{X}/(L_{K}f_{b})$ as that of the Solar neighborhood. However, almost all GCs are located above the magenta strip in Figure~\ref{fig:LX-Gamb}b, which suggests that the evolution of primordial binaries in GCs had been substantially altered by dynamical interactions. MS binaries have either been transformed into X-ray-emitting close binaries, or been dynamically disrupted, leading to a larger value of $L_{X}/(L_{K}f_{b})$ with respect to the field.

\section{Testing the Hills-Heggie Law}

\begin{figure}[ht]
\includegraphics[angle=0,origin=br,height=0.3\textheight, width=0.5\textwidth]{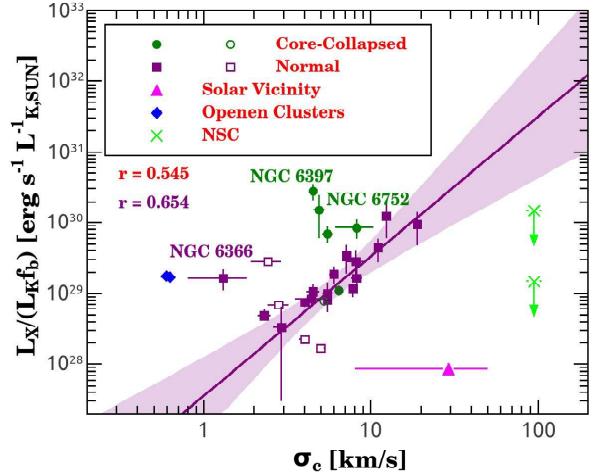}
\linespread{-1.0}
\caption{$L_{X}/(L_{K}f_{b})$ as a function of cluster central velocity dispersion. Color-coded symbols and texts represent the different types of clusters and corresponding corrrelation coefficients as in Figure 1. The solid lines are the best-fitting functions of dynamically normal GCs (GCs with upper limit were not included in the fit), while the shaded area represent the $95\%$ confidence of linear regression. The magenta up-triangle marks the Solar neighborhood stars, which has a large scatter in $\sigma$ among stars of different ages \citep{griv2009}. The blue diamonds represent the two open clusters, with $\sigma=0.62\pm0.1$ $\rm km\,s^{-1}$ for NGC 6791 \citep{tofflemire2014} and $\sigma=0.59^{+0.07}_{-0.06}$ $\rm km\,s^{-1}$ for M 67 \citep{geller2015}. \label{hardness-sigma}}
\end{figure}

In Paper I, we found no significant dependence of the abundance of weak X-ray sources (traced by $L_{X}/L_{K}$) on $f_{b}$ or the cluster central velocity dispersion $\sigma_{c}$. On the other hand, $f_{b}$ was found to be anti-correlated with $\sigma_{c}$ \citep{milone2012}. 
According to the Hills-Heggie law, MS binaries will be gradually exhausted in GCs, and the average stellar kinetic energy plays a vital role in determining whether a MS binary could be dynamically disrupted or be dynamically transformed into X-ray emitting close binaries.
Therefore, with the observed binary hardness ratio of GCs, we can test the Hills-Heggie law as in Figure~\ref{hardness-sigma}, where $L_{X}/(L_{K}f_{b})$ was plotted against $\sigma_{c}$\footnote{We emphasize that here $\sigma_{c}$ is more suitable than $\sigma(r)$ in determining the evolution of binaries in GCs, since binaries tend to sink to the cluster core under the effect of mass segregation, where stellar dynamical interactions are more frequent.}.
The value of $\sigma_{c}$ is adopted from \citep{harris1996}, which has a relative error $\lesssim10\%$ in most GCs.
Clearly, there is a positive correlation between $L_{X}/(L_{K}f_{b})$ and $\sigma_c$ in dynamically normal GCs. The Spearman's rank correlation coefficient and random correlation p-value is $r=0.654$ and $p=0.0018$, respectively. We fit the dynamically normal GCs with a power-law function, which gives $L_{X}/(L_{K}f_{b}) \propto \sigma^{1.71\pm0.48}$ (purple line in Figure~\ref{hardness-sigma}).
We note that the fitted slope is consistent with a value of $2$, which is the case if the binary hardness ratio has a strong dependence on the average stellar kinetic energy ($E_{k}\propto \sigma^{2}$).

As self-gravitating systems, GCs are unstable against core collapse without some source of internal energy. Binaries serve as a reservoir of energy in GCs. The Hills-Heggie law predicts that they will support the clusters from collapse, provided that they can be effectively transformed into harder systems through {\it b-s} and {\it b-b} encounters \citep{hut1983}. 
This prediction can also be tested with the correlation shown in Figure~\ref{hardness-sigma}. 
Note that binaries in core-collapsed GCs exhibit a higher binary hardness ratio than in dynamically normal GCs, which suggests that these systems are running out of their MS binary systems. Indeed, core-collapsed GCs have systematically lower $f_b$ than dynamically normal GCs \citep{milone2012}. The contraction of cluster core will shorten the timescale of binary encounter, thereby boosting the extraction of energy from harder binaries. 

Observationally, the hardening of binaries in GCs can also be tested with the luminosity function (LF, in the form of $dN/ d {\rm log} L_{X} \propto L_{X}^{-\alpha}$) of the weak X-ray sources. 
For example, the dynamically older GCs (with higher central stellar density $\rho_{c}$) were found to host more bright X-ray sources, and the value of $\alpha$ was found to be anti-correlated with $\rho_{c}$ in GCs \citep{pooley2002}. 
Comparing to the Solar neighborhood stars (with $\alpha\approx1.12$, \citet{sazonov2006}), the power-law slopes of GC-LFs were found to be much flatter (with $\alpha<1$), which suggests that dynamical interactions are effective in transforming binaries into bright X-ray sources in GCs. 

For comparison, we also plot the cases of the NSC, Solar neighborhood and two open clusters (i.e., NGC 6791 and M 67) in Figure~\ref{hardness-sigma}. 
The NSC and Solar neighborhood were shown as green crosses and magenta up-triangle separately; the two open clusters, with $L_{X}/(L_{K}f_{b})=(1.7\pm0.2)\times 10^{29} \rm \ erg\,s^{-1} M^{-1}_{\odot}$ for NGC 6791 and $L_{X}/(L_{K}f_{b})=1.76 \times 10^{29} \rm \ erg\,s^{-1} M^{-1}_{\odot}$ for M 67, were plotted as blue diamonds. 
To obtain these values, we have adopted the X-ray emissivity ($L_{X}/L_{K}$) from \citet{vandenberg2013}, and the MS binary fraction of $f_{b}=30\pm5\%$ for NGC 6791 \citep{bedin2008} and $f_{b}=45\%$ for M 67 \citep{davenport2010}, respectively. 

Due to the collisionless environment, although Solar neighborhood stars show larger $f_{b}$ and $\sigma$ than the clusters, their binary hardness ratio is the lowest in Figure~\ref{hardness-sigma}.
For open clusters, they have larger binary hardness ratio than the Solar neighborhood and some of the GCs, but these systems tend to disperse quickly, hence evaporation and mass segregation effects will leave these systems with a large fraction of MS binaries (with $f_{b}$ comparable to the Solar neighborhood) and the lowest stellar velocity dispersion. Hence the higher value of $L_{X}/(L_{K}f_{b})$ in open clusters is more likely due to the substantial loss of single stars rather than dynamical hardening of MS binaries. 
In fact, the location of open clusters in Figure~\ref{hardness-sigma} is close to NGC\,6366, a GC known to suffer from strong tidal stripping \citep{paust2009}. 
Notably, the NSC shows a much lower binary hardness ratio than that predicted by the GCs in Figure~\ref{hardness-sigma}. The NSCS also exhibits a steep LF, with $\alpha\approx 1.63$ over the luminosity range of $10^{31-33}\, \rm erg\,s^{-1}$ \citep{zhu2018}.
The implication of these findings will be addressed in Section 5. 

\section{Discussion: NSC in Context}

As the most massive and the densest star cluster in the Milky Way, the NSC offers a unique laboratory for studying stellar dynamics in high stellar velocity dispersion environment, especially under the gravitational influence of the super-massive black hole (SMBH). 
The X-ray sources detected in the NSC, mainly close binaries with accreting compact objects, may serve as sensitive probes of the stellar dynamics in this dense environment.
It is also important to make connection between the NSC and GCs, as the former has been suggested to be assembled, at least in part, by sequential mergers of GCs that spiraled into the deep gravitational well of the Galactic center, due to dynamical friction \citep{tremaine1975,antonini2012,antonini2013,arca2014,gnedin2014}.
Alternatively, the NSC might be formed through continuous {\it in-situ} star formation, supplied by gas inflow that is driven by some still poorly understood processes \citep{milosavljevi2004,emsellem2008}.

In Figure~1, we found that the NSC is compatible with the correlations defined by the GCs, which may imply for a common origin for the weak X-ray sources in GCs and the NSC.
Indeed, there are many similarities between NSC and GCs in the populations of X-ray sources. 
For example, about a dozen bright transient X-ray sources, mainly BH-LMXBs and NS-LMXBs, have been detected in the NSC \citep{muno2005,degenaar2015}.
The spatial distribution of these transient sources was found to be scaled with the square of the stellar density profile ($\rho^{2}(r)$) of the NSC, strongly suggesting a dynamical origin \citep{zhu2018}.
On the other hand, the weak and steady X-ray sources, i.e., CVs, show a spatial distribution that matches well with the stellar density profile ($\rho(r)$; \citealp{zhu2018}). At first sight, these findings may suggest a universal scenario for forming X-ray sources in dense stellar environments including GCs and the NSC: a dynamical origin is responsible for the over-abundance of NS-LMXBs or BH-LMXBs, while CVs are mainly descendant from the primordial binaries that have been modified by stellar dynamical interactions.

Nevertheless, the stellar velocity dispersion in the NSC, $\sigma\sim 100\,\rm km\,\rm s^{-1}$, is about one order of magnitude higher than that in GCs, which suggests a much shorter watershed orbital period ($P_{w}$) for binaries in the NSC according to the Hills-Heegie law. 
Furthermore, the stars used to quantify $\sigma$ in the NSC are typically younger and more massive (with an average mass of $\sim 1.5-3M_{\odot}$; \citealp{schodel2007}) than the older stellar populations represented by the X-ray sources. With these conditions taken into account, we estimate $P_{w}\sim 1-4\, \rm days$ for binaries with a stellar mass of $1{\rm~M_{\odot}}$ in the NSC. Such a watershed period is much shorter than the typical period of ABs ($P_{orb}\sim 10$ days; \citealp{eker2008}), indicating that even close binaries like ABs could be disrupted dynamically in the NSC.
On the other hand, $P_{w}$ is comparable to the maximum period of CVs and LMXBs \citep[7.23 edition]{ritter2003}, thus formation of these systems through {\it b-s}/{\it b-b} encounters are inefficient in the NSC.

For the BH-LMXBs and NS-LMXBs in the NSC, \citet{generozov2018} suggested that they could be formed by tidal capture of stars by BHs adn NSs. Alternatively, NS-LMXBs and their descendant, MSPs, could be inherited from GCs having inspiralled into the Galactic center \citep{arca2018}.
For the CVs, their abundance was found to be slightly higher in the NSC with respect to the field CV population over the same luminosity range \citep{zhu2018}. According to the Hills-Heggie law, this feature is inconsistent with the in-situ star formation scenario of the NSC, since most of the primordial binaries would have been disrupted dynamically before they can otherwise evolve into CVs. 
However, if most CVs in the NSC are inherited from the dense cores of GCs, their present-day abundance would depend on the evolutionary phase of the parent GCs \citep{arca2018}. The dynamically older GCs tend to have a high abundance of weak X-ray sources (Paper I), and the host galaxy's tidal field will also accelerate the evolution of GCs \citep{gnedin1999}.

The NSC was found to show a much lower $L_{X}/(L_{K}f_{b})$ than predicted for its $\sigma$ (Figure~\ref{hardness-sigma}). We suggest that this discrepancy may be caused by the different population of X-ray sources in GCs and NSC, which is ultimately related to the difference in the watershed period (velocity) that distinguishes soft/hard binaries, i.e., $P_{w}\approx 5000$ days in GCs versus $P_{w}\approx 1-4$ days in the NSC. 
As discussed in the above, the small-$P_{w}$ environment of the NSC disfavors the formation of ABs, and in the meantime tends to accelerate the evolution of CVs, driving them towards smaller $P_{\rm orb}$, lower mass transfer rates and lower X-ray luminousities \citep{patterson1984,townsley2009,patterson2011}.    
This in turn results in a lower binary hardness ratio in the NSC.

Alternatively, the low $L_{X}/(L_{K}f_{b})$ in the NSC might be partially explained by the presence of the SMBH, also known as Sgr $\rm A^{\ast}$. The NSC stars are expected to be heated when they come close to the SMBH, as a result, the hardening of binaries is inefficient and binaries are more likely to be disrupted. Such an effect has been predicted by the simulations of \citet{hopman2009}. 

\section{Conclusion}

To bring isolated stars (or loosely bounded binaries) together to form X-ray sources, the fundamental issue is how to dissipate the stars' kinetic energy (or binary bounding energy) effectively. Early studies of stellar dynamical interactions in dense stellar environments have provided various scenarios for solving this problem. 
In this work, we strengthen the importance of {\it b-s} and {\it b-b} encounters as an effective formation mechanism for X-ray sources in GCs. 
We have demonstrated a tight correlation between the MS binary encounter rate ($\Gamma_{b}$) and the total luminosity (hence number) of weak X-ray sources in 30 Galactic GCs. 
Using a physical parameter $L_{X}/(L_{K}f_{b})$ for these GCs, we have verified the Hills-Heggie law, which states that stellar encounters involving hard binaries make them harder, whereas encounters involving soft binaries drive them softer and eventually disrupted.
Applying the Hills-Heggie law to the dynamic environment of the NSC, we argue that both the dynamical and primordial channels of CV formation are suppressed,
and that a large fraction of the weak X-ray sources detected therein might have been inherited from GCs captured into the Galactic center.

\acknowledgements
We thank the anonymous referee for valuable comments that help improve our manuscript. This work is supported by the National Key R\&D Program of China No. 2017YFA0402600, the National Science Fundation of China under grants 11525312, 11133001, 11333004 and 11303015.
 
\begin{deluxetable}{rrrrrrrrr}
\tabletypesize{\scriptsize}
\tablecolumns{9}
\linespread{1.1}
\tablewidth{0pc}
\tablenum{1}
\tablecaption{Basic Properties of Star Clusters}
\tablehead{
\colhead{Name} & \colhead{$\Gamma$} & \colhead{$f_{b}$} & \colhead{$\Gamma_{b}$} & \colhead{$\gamma$} & \colhead{$\sigma$} & \colhead{$L_{X}$} & \colhead{$L_{K}$} & \colhead{$L_{X}/(L_{K}f_{b}$)} \\
\colhead{---} & \colhead{---} & \colhead{$\%$} & \colhead{---} & \colhead{---} & \colhead{$km\,s^{-1}$} & \colhead{$10^{32}\rm\ erg\,s^{-1}$} & \colhead{$10^{4}L_{K,\odot}$} & \colhead{$10^{28}\rm\ erg\,s^{-1}L_{K,\odot}^{-1}$}\\
\colhead{(1)} & \colhead{(2)} & \colhead{(3)} & \colhead{(4)} & \colhead{(5)} & \colhead{(6)} & \colhead{(7)} & \colhead{(8)} & \colhead{(9)}}
\startdata
\sidehead{Normal GCs:}
NGC\,104  & $1000^{+150}_{-130}$ & $1.8 \pm0.6$ & $180^{+66}_{-65}$      & $840^{+130}_{-110}$   & $11.0\pm0.3$& $58.9^{+7.7}_{-8.6}$ & 74.6  & $44^{+16}_{-16}$\\
NGC\,288  &$0.77^{+0.28}_{-0.21}$& $11.2\pm0.8$ & $0.86^{+0.32}_{-0.24}$ & $7.6^{+2.8}_{-2.0}$   & $2.9\pm0.3$ & $1.3^{+1.2}_{-1.2}$  & 3.4   & $3.3^{+3.1}_{-3.0}$\\
NGC\,3201 & $7.2^{+3.6}_{-2.3}$  & $12.2\pm0.6$ & $8.8^{+4.4}_{-2.8}$    & $37^{+18}_{-12}$      & $5.0\pm0.2$ & $<1.5           $    & 7.2   & $<1.7$ \\
NGC\,5024 & $35^{+12}_{-10}$     & $6.2\pm0.6$  & $22.0^{+8.0}_{-6.3}$   & $58^{+20}_{-16}$      & $4.4\pm0.9$ & $12.1^{+2.6}_{-2.6}$ & 23.1  & $8.5^{+2.0}_{-2.0}$\\
NGC\,5272 & $194^{+33}_{-18}$    & $3.4\pm0.6$  & $66^{+16}_{-13}$       & $269^{+46}_{-25}$     & $5.5\pm0.3$ & $9.1^{+3.1}_{-2.6}$  & 33.0  & $8.2^{+3.1}_{-2.8}$\\ 
NGC\,5286 & $458^{+58}_{-61}$    & $1.8\pm0.6$  & $82^{+29}_{-30}$       & $723^{+92}_{-96}$     & $8.1\pm0.1$ & $13.5^{+3.6}_{-2.6}$ & 26.1  & $29^{+12}_{-11}$\\
NGC\,5904 & $164^{+39}_{-30}$    & $2.2\pm0.6$  & $36^{+13}_{-12}$       & $243^{+57}_{-45}$     & $5.5\pm0.4$ & $5.1^{+1.4}_{-0.8}$  & 22.8  & $10.1^{+3.9}_{-3.2}$\\
NGC\,5927 & $68^{+13}_{-10}$     & $3.2\pm0.6$  & $21.8^{+5.8}_{-5.3}$   & $254^{+47}_{-38}$     & ...         & $4.9^{+1.2}_{-0.8}$  & 17.3  & $8.8^{+2.8}_{-2.2}$\\
NGC\,6093 & $532^{+59}_{-69}$    & $1.2\pm0.6$  & $64^{+33}_{-33}$       & $1340^{+150}_{-170}$  & $12.4\pm0.6$& $23.7^{+1.4}_{-1.8}$ & 16.1  & $123^{+62}_{-62} $\\ 
NGC\,6121 & $27^{+12}_{-10}$     & $12.2\pm0.8$ & $33^{+14}_{-12}$       & $177^{+76}_{-63}$     & $4.0\pm0.2$ & $5.3^{+0.0}_{-0.7}$  & 5.8   & $7.5^{+0.5}_{-1.1}$\\
NGC\,6144 & $3.1^{+1.1}_{-0.9}$  & $9.6\pm0.6$  & $3.0^{+1.0}_{-0.8}$    & $28.3^{+9.6}_{-7.7}$  & ...         & $5.0^{+1.4}_{-0.8}$  & 4.7   & $11.0^{+3.1}_{-1.9}$\\
NGC\,6205 & $69^{+18}_{-15}$     & $1.4\pm0.6$  & $9.7^{+4.9}_{-4.6}$    & $130^{+34}_{-28}$     & $7.1\pm0.4$ & $9.2^{+0.6}_{-0.7}$  & 19.4  & $34^{+15}_{-15}$\\
NGC\,6218 & $13.0^{+5.4}_{-4.0}$ & $7.4\pm0.6$  & $9.6^{+4.1}_{-3.1}$    & $77^{+32}_{-24}$      & $4.5\pm0.4$ & $4.0^{+1.0}_{-0.8}$  & 5.1   & $10.6^{+2.7}_{-2.3}$\\
NGC\,6341 & $270^{+30}_{-29}$    & $2.2\pm0.6$  & $59^{+18}_{-17}$       & $695^{+77}_{-75}$     & $6.0\pm0.4$ & $5.4^{+0.5}_{-0.5}$  & 13.2  & $18.6^{+5.4}_{-5.3}$\\
NGC\,6352 & $6.7^{+1.7}_{-1.3}$  & $13.8\pm0.8$ & $9.3^{+2.4}_{-1.9}$    & $86^{+22}_{-17}$      & ...         & $4.1^{+0.8}_{-0.9}$  & 6.1   & $4.9^{+1.0}_{-1.1}$\\
NGC\,6362 & $4.6^{+1.5}_{-1.0}$  & $9.2\pm0.6$  & $4.2^{+1.4}_{-1.0}$    & $37^{+12}_{-9}$       & $2.8\pm0.4$ & $<3.9         $      & 6.1   & $<6.9$\\
NGC\,6366 & $5.1^{+2.8}_{-1.8}$  & $18.4\pm1.4$ & $9.5^{+5.1}_{-3.3}$    & $129^{+69}_{-44}$     & $1.3\pm0.5$ & $3.2^{+1.0}_{-1.0}$  & 1.1   & $16.3^{+5.5}_{-5.2}$\\
NGC\,6388 & $900^{+240}_{-210}$  & $1.6\pm0.8$  & $144^{+81}_{-80}$      & $770^{+200}_{-180}$   & $18.9\pm0.8$& $88.8^{+1.7}_{-2.8}$ & 58.4  & $95^{+48}_{-48}$\\
NGC\,6535 &$0.39^{+0.40}_{-0.20}$& $6.6\pm1.8$  & $0.26^{+0.27}_{-0.14}$ & $24^{+24}_{-12}$      & $2.4\pm0.5$ & $<1.2             $  & 0.6   & $<28.7$\\
NGC\,6637 & $90^{+36}_{-18}$     & $6.0\pm0.6$  & $54^{+22}_{-12}$       & $390^{+160}_{-80}$    & ...         & $16.6^{+5.6}_{-3.7}$ & 16.0  & $17.3^{+6.1}_{-4.3}$\\
NGC\,6656 & $78^{+32}_{-26}$     & $4.4\pm0.6$  & $34^{+15}_{-12}$       & $153^{+63}_{-51}$     & $7.8\pm0.3$ & $10.6^{+0.3}_{-2.1}$ & 20.8  & $11.6^{+1.6}_{-2.7}$\\
NGC\,6809 & $3.2^{+1.4}_{-1.0}$  & $8.0\pm0.6$  & $2.6^{+1.1}_{-0.8}$    & $15.0^{+6.4}_{-4.6}$  & $4.0\pm0.3$ & $<1.3              $ & 7.0   & $<2.2$\\
NGC\,6838 & $1.5^{+0.2}_{-0.1}$  & $23.4\pm1.4$ & $3.4^{+0.4}_{-0.4}$    & $41.5^{+4.1}_{-3.9}$  & $2.3\pm0.2$ & $2.7^{+0.5}_{-0.4}$  & 2.3   & $4.9^{+1.0}_{-0.7}$\\
NGC\,7089 & $518^{+78}_{-71}$    & $2.6\pm0.6$  & $135^{+37}_{-36}$      & $626^{+94}_{-86}$     & $8.2\pm0.6$ & $14.2^{+6.4}_{-4.0}$ & 33.2  & $16.5^{+8.3}_{-6.0}$\\

\cline{1-9}
\sidehead{Core-Collapsed GCs:}
NGC\,362  & $740^{+140}_{-120}$  & $4.0\pm0.6$  & $294^{+70}_{-64}$      & $1540^{+290}_{-250}$  & $6.4\pm0.3$ & $9.2^{+0.3}_{-0.5}$  & 20.6  & $11.1^{+1.7}_{-1.8}$\\
NGC\,6397 & $840^{+180}_{-180}$  & $2.4\pm0.6$  & $20.2^{+6.7}_{-6.7}$   & $920^{+200}_{-200}$   & $4.5\pm0.2$ & $13.3^{+0.5}_{-0.5}$ & 2.0   & $281^{+71}_{-71}$\\
NGC\,6541 & $386^{+95}_{-63}$    & $2.0\pm0.6$  & $77^{+30}_{-26}$       & $750^{+180}_{-120}$   & $8.2\pm2.1$ & $22.6^{+0.1}_{-0.1}$ & 13.3  & $85^{+26}_{-26}$\\
NGC\,6681 & $1040^{+270}_{-190}$ & $3.8\pm0.6$  & $400^{+120}_{-100}$    & $7300^{+1800}_{-1400}$& $5.2\pm0.5$ & $<1.4             $  & 4.6   & $<7.9$\\
NGC\,6752 & $400^{+180}_{-130}$  & $1.0\pm0.6$  & $40^{+30}_{-27}$       & $1610^{+730}_{-500}$  & $4.9\pm0.4$ & $11.1^{+0.4}_{-0.4}$ & 7.3   & $153^{+92}_{-92}$\\  
NGC\,7099 & $320^{+120}_{-80}$   & $2.4\pm0.6$  & $78^{+36}_{-28}$       & $1680^{+640}_{-420}$  & $5.5\pm0.4$ & $8.7^{+0.6}_{-0.6}$  & 5.2   & $70^{+18}_{-18}$\\
\cline{1-9}
\sidehead{Galactic Nuclear Star Cluster (with parameters obtained within $r\leqslant 100\arcsec$):}
NSC    & $25600^{+7700}_{-7700}$ & $1.0-10.0$  & $2560-25600$ & $4200^{+1200}_{-1200}$  & $90-100$ & $<1374$ & $1194\pm339 $    & $<14.8-148$\\ 
\enddata
\vspace{-0.5cm}
\tablecomments{(1) Target name; (2) Encounter rate adopted from \citet{bahramian2013}, for the NSC, the encounter rates are estimated in this work (see Section 3); (3) Main sequence binary fraction in units of percentage \citep{milone2012}, the binary fraction of the NSC is estimated in this work (see Section 3); (4) Binary-single and binary-binary encounter rate; (5) The specific encounter rate; (6) Cluster velocity dispersion in units of $\rm km\,s^{-1}$ \citep{harris1996}; (7) 0.5-8 keV cumulative luminosity in units of $10^{32}\rm\ erg\,s^{-1}$; (8) K-band luminosity in units of $10^{4}L_{K,\odot}$. For the NSC, a cumulative mass of $(609\pm97)\times 10^{6}\, \rm M_{\odot}$ within $r\leq 100\arcsec$ is converted into the K-band luminosity using a mass-to-light ratio of $M/L_K=0.51\pm0.12M_{\odot}/L_{K,\odot}$ \citep{fritz2016}; (9) Binary hardness ratio, in units of $10^{28}\rm\ erg\,s^{-1}L_{K,\odot}^{-1}$.}
\end{deluxetable}

\begin{acknowledgements}

\end{acknowledgements}

%\newpage
%\clearpage
\label{lastpage}


\begin{thebibliography}
\expandafter\ifx\csname natexlab\endcsname\relax\def\natexlab#1{#1}\fi
\bibitem[Antonini et al.(2012)]{antonini2012} Antonini, F., Capuzzo-Dolcetta, R., Mastrobuono-Battisti, A., \& Merritt, D.\ 2012, \apj, 750, 111 
\bibitem[Antonini(2013)]{antonini2013} Antonini, F.\ 2013, \apj, 763, 62 
\bibitem[Arca-Sedda \& Capuzzo-Dolcetta(2014)]{arca2014} Arca-Sedda, M., \& Capuzzo-Dolcetta, R.\ 2014, \mnras, 444, 3738 
\bibitem[Arca-Sedda et al.(2018)]{arca2018} Arca-Sedda, M., Kocsis, B., \& Brandt, T.~D.\ 2018, \mnras,  
\bibitem[Bacon et al.(1996)]{bacon1996} Bacon, D., Sigurdsson, S., Davis, M. B., 1996, \mnras, 281, 830
\bibitem[Bahramian et al.(2013)]{bahramian2013} Bahramian, A., Heinke, C.O., Sivakoff, G.R., \& Gladstone, J.C. 2013, \apj, 766, 136
\bibitem[Bedin et al.(2008)]{bedin2008} Bedin, L.~R., Salaris, M., Piotto, G., et al.\ 2008, \apjl, 679, L29 
\bibitem[Chatterjee et al.(2013)]{chatterjee2013} Chatterjee, S., Rasio, F.~A., Sills, A., \& Glebbeek, E.\ 2013, \apj, 777, 106 
\bibitem[Cheng et al.(2018)]{cheng2018} Cheng, Z., Li, Z., Xu, X., \& Li, X.\ 2018, \apj, 858, 33 
\bibitem[Clark(1975)]{clark1975} Clark, G. W. 1975, \apj, 199, L143
%\bibitem[Cohn et al.(2010)]{cohn2010} Cohn, H.~N., Lugger, P.~M., Couch, S.~M., et al.\ 2010, \apj, 722, 20 
\bibitem[Davies(2002)]{davies2002} Davies, M. B., 2002, ASPC, 263, 17
\bibitem[Davenport et al.(2010)]{davenport2010} Davenport, J. R. A., \& Sandquist, E. L., 2010, \apj, 711, 559
\bibitem[Degenaar et al.(2015)]{degenaar2015} Degenaar, N., Wijnands, R., Miller, J.~M., et al.\ 2015, Journal of High Energy Astrophysics, 7, 137 
\bibitem[de Grijs et al.(2013)]{degrijs2013} de Grijs, R., Li, C., Zheng, Y., et al.\ 2013, \apj, 765, 4 
\bibitem[Duquennoy \& Mayor(1991)]{duquennoy1991} Duquennoy, A., Mayor, M., 1991, \aa, 248, 485
\bibitem[Eker et al.(2008)]{eker2008} Eker, Z., Ak, N.~F., Bilir, S., et al.\ 2008, \mnras, 389, 1722 
\bibitem[Emsellem \& van de Ven(2008)]{emsellem2008} Emsellem, E., \& van de Ven, G.\ 2008, \apj, 674, 653 
\bibitem[Fabian et al.(1975)]{fabian1975} Fabian, A. C., Pringle, J. E., \& Rees, M. J. 1975, \mnras, 172, 15P
\bibitem[Fregeau et al.(2004)]{fregeau2004} Fregeau, J.~M., Cheung, P., Portegies Zwart, S.~F., \& Rasio, F.~A.\ 2004, \mnras, 352, 1 
\bibitem[Fritz et al.(2016)]{fritz2016} Fritz, T.~K., Chatzopoulos, S., Gerhard, O., et al.\ 2016, \apj, 821, 44 
\bibitem[Geller et al.(2015)]{geller2015} Geller, A. M., Latham, D. W., Mathieu, R. D., 2015, \apj, 150, 97
\bibitem[Generozov et al.(2018)]{generozov2018} Generozov, A., Stone, N.~C., Metzger, B.~D., \& Ostriker, J.~P.\ 2018, arXiv:1804.01543
\bibitem[Gnedin et al.(1999)]{gnedin1999} Gnedin, O.~Y., Lee, H.~M., \& Ostriker, J.~P.\ 1999, \apj, 522, 935 
\bibitem[Gnedin et al.(2014)]{gnedin2014} Gnedin, O.~Y., Ostriker, J.~P., \& Tremaine, S.\ 2014, \apj, 785, 71  
\bibitem[Griv et al.(2009)]{griv2009} Griv, E., Gedalin, M., \& Eichler, D. 2009, \aj, 137, 3520
\bibitem[G{\"u}del(2004)]{gudel2004} G{\"u}del, M.\ 2004, \aapr, 12, 71 
\bibitem[Hailey et al.(2016)]{hailey2016} Hailey, C.~J., Mori, K., Perez, K., et al.\ 2016, \apj, 826, 160 
\bibitem[Hailey et al.(2018)]{hailey2018} Hailey, C.~J., Mori, K., Bauer, F.~E., et al.\ 2018, \nat, 556, 70 
\bibitem[Harris(2010 edition)]{harris1996} Harris, W. E. 1996(2010 edition), \aj, 112, 1487.
\bibitem[Heggie(1975)]{heggie1975} Heggie, D. C., 1975, \mnras, 173, 729
\bibitem[Heggie et al.(1996)]{heggie1996} Heggie, D.~C., Hut, P., \& McMillan, S.~L.~W.\ 1996, \apj, 467, 359 
\bibitem[Heinke(2010)]{heinke2010} Heinke, C. O. 2010, in AIP Conf. Proc. 1314, International Conference of Binaries: In celebration of Ron Webbink's 65th Birthday (Melville, NY: AIP), 135
\bibitem[Hills(1975)]{hills1975} Hills, J. G., 1975, \aj, 80, 809
\bibitem[Hills(1976)]{hills1976} Hills, J. G., 1976, \mnras, 175, 1P
\bibitem[Hoffer(1983)]{hoffer1983} Hoffer, J. B., 1983, \aj, 88, 1420
\bibitem[Hopman(2009)]{hopman2009} Hopman, C.\ 2009, \apj, 700, 1933 
\bibitem[Hut(1983)]{hut1983} Hut, P., 1983, \apj, 272, L29
\bibitem[Hut et al.(1992a)]{hut1992a} Hut, P., McMillan, S., Romani, R. W., 1992, \apj, 389, 527
\bibitem[Hut et al.(1992b)]{hut1992b} Hut, P., McMillan, S., Goodman, J., et al. 1992, \pasp, 104, 981
\bibitem[Hut(1993)]{hut1993} Hut, P., 1993, \apj, 403, 256
\bibitem[Ivanova et al.(2006)]{ivanova2006} Ivanova, N., Heinke, C. O., Rasio, F. A., Taam, R. E., et al. 2006, \mnras, 372, 1043
\bibitem[Ivanova et al.(2008)]{ivanova2008} Ivanova, N., Heinke, C. O., Rasio, F. A., et al. 2008, \mnras, 386, 553
\bibitem[Ji \& Bregman(2015)]{ji2015} Ji, J., Bregman, J. N., 2015, \apj, 807, 32
\bibitem[Jord$\acute{a}$n et al.(2004)]{jordan2004} Jord$\acute{a}$n, A., et al. 2004, \apj, 613, 279
\bibitem[Jord$\acute{a}$n et al.(2007)]{jordan2007} Jord$\acute{a}$n, A., et al. 2007, \apjl, 671, 117
\bibitem[Just et al.(2015)]{just2015} Just, A., Fuchs, B., Jahrei{\ss}, H., et al.\ 2015, \mnras, 451, 149 
\bibitem[Katz(1975)]{katz1975} Katz, J. I. 1975, \nat, 253, 698
%\bibitem[Knigge et al.(2011)]{knigge2011} Knigge, C., Baraffe, I., \& Patterson, J.\ 2011, \apjs, 194, 28 
%\bibitem[Lugger et al.(2017)]{lugger2017} Lugger, P.~M., Cohn, H.~N., Cool, A.~M., Heinke, C.~O., \& Anderson, J.\ 2017, \apj, 841, 53 
\bibitem[Maxwell et al.(2012)]{maxwell2012} Maxwell, J. E., Lugger, P. M., Cohn, H. N., et al. 2012, \apj, 756, 147 
\bibitem[Muno et al.(2005)]{muno2005} Muno, M.~P., Pfahl, E., Baganoff, F.~K., et al.\ 2005, \apjl, 622, L113 
\bibitem[Mikkola(1983)]{mikkola1983} Mikkola, S., 1983, \mnras, 203, 1107
\bibitem[Mikkola(1984a)]{mikkola1984a} Mikkola, S., 1984, \mnras, 207, 115
\bibitem[Mikkola(1984b)]{mikkola1984b} Mikkola, S., 1984, \mnras, 208, 75
\bibitem[Milone et al.(2012)]{milone2012} Milone, A. P., Piotto, G., Bedin, L. R., et al. 2012, \aap, 540, 16
\bibitem[Milosavljevi{\'c}(2004)]{milosavljevi2004} Milosavljevi{\'c}, M.\ 2004, \apjl, 605, L13 
%\bibitem[Panamarev et al.(2018)]{panamarev2018} Panamarev, T., Just, A., Spurzem, R., et al.\ 2018, arXiv:1805.02153 
\bibitem[Patterson(1984)]{patterson1984} Patterson, J.\ 1984, \apjs, 54, 443 
\bibitem[Patterson(2011)]{patterson2011} Patterson, J.\ 2011, \mnras, 411, 2695 
\bibitem[Paust et al.(2009)]{paust2009} Paust, N. E. Q., Aparicio, A., Piotto, G., et al. 2009, \aj, 137, 246
\bibitem[Peacock et al.(2009)]{peacock2009} Peacock, M. B., et al. 2009, \mnras, 392, 55
\bibitem[Pooley et al.(2002)]{pooley2002} Pooley, D., Lewin, W. H. G., Verbunt F., et al. 2002, \apj, 573, 184
\bibitem[Pooley et al.(2003)]{pooley2003} Pooley, D., Lewin, W.~H.~G., Anderson, S.~F., et al.\ 2003, \apjl, 591, L131
\bibitem[Pooley \& Hut(2006)]{pooley2006} Pooley, D., \& Hut, P. 2006, \apj, 646, 143
\bibitem[Raghavan et al.(2010)]{raghavan2010} Raghavan, D., McAlister, H. A., Henry, T. J., et al. 2010, \apjs, 190, 1
\bibitem[Rasio et al.(2000)]{rasio2000} Rasio, F.~A., Pfahl, E.~D., \& Rappaport, S.\ 2000, \apjl, 532, L47 
\bibitem[Reis et al.(2013)]{reis2013} Reis, R.~C., Wheatley, P.~J., G{\"a}nsicke, B.~T., \& Osborne, J.~P.\ 2013, \mnras, 430, 1994 
\bibitem[Revnivtsev et al.(2007)]{revnivtsev2007} Revnivtsev, M., Churazov, E., Sazonov, S., Forman, W., \& Jones,C. 2007, \aap, 473, 783
\bibitem[Ritter \& Kolb(2003)]{ritter2003} Ritter, H., \& Kolb, U.\ 2003, \aap, 404, 301 
%\bibitem[Rivera Sandoval et al.(2018)]{sandoval2018} Rivera Sandoval, L.~E., van den Berg, M., Heinke, C.~O., et al.\ 2018, \mnras, 475, 4841 
\bibitem[Sazonov et al.(2006)]{sazonov2006} Sazonov, S., Revnivtsev, M., Gilfanov, M., et al. 2006, \aap, 450, 117
\bibitem[Sch{\"o}del et al.(2007)]{schodel2007} Sch{\"o}del, R., Eckart, A., Alexander, T., et al.\ 2007, \aap, 469, 125 
\bibitem[Sivakoff et al.(2007)]{sivakoff2007} Sivakoff, G. R., Jordan, A., Sarazin, C. L., et al. 2007, \apj, 660, 1246
\bibitem[Sollima et al.(2007)]{sollima2007} Sollima, A., Beccari, G., Ferraro, F. R., et al. 2007, \mnras, 380, 781
\bibitem[Sutantyo(1975)]{sutantyo1975} Sutantyo, W. 1975, \aap, 44, 227
\bibitem[Tremaine et al.(1975)]{tremaine1975} Tremaine, S.~D., Ostriker, J.~P., \& Spitzer, L., Jr.\ 1975, \apj, 196, 407 
\bibitem[Tofflemire et al.(2014)]{tofflemire2014} Tofflemire, B.M., Gosnell, N. M., et al. 2014, \aj 148, 61
\bibitem[Townsley \& G{\"a}nsicke(2009)]{townsley2009} Townsley, D.~M., \& G{\"a}nsicke, B.~T.\ 2009, \apj, 693, 1007 
\bibitem[van den Berg et al.(2013)]{vandenberg2013} van den Berg, M., Verbunt, F., Tagliaferri, G., et al. 2013, \apj, 770, 98
\bibitem[Verbunt \& Hut(1987)]{verbunt1987} Verbunt, F., \& Hut, P. 1987, in: The Origin and Evolution of Neutron Stars IAU Symp.125, eds. D.J. Helfand and J.H. Huang, Reidel, p.187
\bibitem[Verbunt(2003)]{verbunt2003} Verbunt, F.\ 2003, New Horizons in Globular Cluster Astronomy, 296, 245 
\bibitem[Zhu et al.(2018)]{zhu2018} Zhu, Z., Li, Z., \& Morris, M.~R.\ 2018, \apjs, 235, 26 
\bibitem[Xu \& Li(2018)]{xu2018} Xu, X.-j., \& Li, Z.\ 2018, \apj, 856, 77 
\end{thebibliography}
\end{document}